\documentclass[12pt]{article}
\usepackage{subeqnarray,indent,amsfonts,amssymb}
\usepackage{bm}    
\usepackage{graphicx}
\usepackage{url}
\usepackage{hanging}  
\usepackage{epstopdf}

\def\doublespace{ \renewcommand{\baselinestretch}{1.325} \large\normalsize }
\def\singlespace{ \renewcommand{\baselinestretch}{1} \large\normalsize }

\begin{document}

\bibliographystyle{plain}

\date{March 28, 2014}

\title{\vspace*{-2cm}The Persistence of Wishful Thinking: \\[2mm]
       Response to ``Updated Thinking on Positivity Ratios''
          \\[5mm]}

\author{
   {\small Nicholas J.~L.~Brown}    \\[-1mm]
   {\small\it New School of Psychotherapy and Counselling}   \\[4mm]
   {\small Alan D.~Sokal}       \\[-1mm]
   {\small\it New York University and University College London}     \\[4mm]
   {\small Harris L.~Friedman}   \\[-1mm]
   {\small\it Saybrook University and University of Florida}
\vspace*{1cm}
}

\maketitle
\thispagestyle{empty}   

\singlespace
\begin{center}
\begin{quote}
\begin{quote}
A slightly abridged version of this article was published in
\hbox{\em American} {\em Psychologist}\/ {\bf 69}, 629--632 (2014),
\url{http://dx.doi.org/10.1037/a0037050}
\qquad
\copyright\  2014 American Psychological Association

\bigskip

This article may not exactly replicate the final version published in
the APA journal. It is not the copy of record.
\end{quote}
\end{quote}
\end{center}

\bigskip
\noindent
{\bf Running Head:} Response on positivity ratios

\doublespace
\vspace*{1cm}
\begin{center}
{\bf Author note}
\end{center}
\vspace*{-3mm}
\noindent
Nicholas J.~L.~Brown, New School of Psychotherapy and Counselling, London, UK;
\\
Alan~D.~Sokal, Department of Physics, New York University
and Department of Mathematics, University College London;
\\
Harris L.~Friedman,
School of Psychology and Interdisciplinary Inquiry, Saybrook University and
Department of Psychology, University of Florida.

\clearpage

\doublespace
\vspace*{0mm}
\begin{center}
{\bf Abstract}
\end{center}
\vspace*{-3mm}
\noindent
We analyze critically the
renewed
claims made by Fredrickson (2013)
concerning positivity ratios and ``flourishing,''
and attempt to disentangle some conceptual confusions;
we also address the alleged empirical evidence for nonlinear effects.
We conclude that there is no evidence
whatsoever
for the existence of any ``tipping points,''
and only
weak evidence for the existence of any nonlinearity of any kind.
Our original concern, that the application of advanced mathematical techniques
in psychology and related disciplines may not always be appropriate, remains undiminished.

\bigskip
\bigskip

\clearpage

\newcommand{\apasection}[1]{{\begin{center} {\bf #1} \end{center}\vspace{-3mm}}}
\newcommand{\apasubsection}[1]{{\medskip\par\noindent {\bf #1} \par}}
\newcommand{\apasubsubsection}[1]{{\par\indent {\bf #1.}}}
\newcommand{\apaparagraph}[1]{{\par\indent \textit{\textbf{#1.}} }}

\newcommand{\be}{\begin{equation}}
\newcommand{\ee}{\end{equation}}
\newcommand{\<}{\langle}
\renewcommand{\>}{\rangle}
\newcommand{\widebar}{\overline}
\def\reff#1{(\protect\ref{#1})}
\def\spose#1{\hbox to 0pt{#1\hss}}
\def\ltapprox{\mathrel{\spose{\lower 3pt\hbox{$\mathchar"218$}}
 \raise 2.0pt\hbox{$\mathchar"13C$}}}
\def\gtapprox{\mathrel{\spose{\lower 3pt\hbox{$\mathchar"218$}}
 \raise 2.0pt\hbox{$\mathchar"13E$}}}
\def\textprime{${}^\prime$}

\def\scra{\mathcal{A}}
\def\scrb{\mathcal{B}}
\def\scrc{\mathcal{C}}
\def\scrd{\mathcal{D}}
\def\scre{\mathcal{E}}
\def\scrf{\mathcal{F}}
\def\scrg{\mathcal{G}}
\def\scrl{\mathcal{L}}
\def\scrm{\mathcal{M}}
\def\scro{\mathcal{O}}
\def\scrp{\mathcal{P}}
\def\scrq{\mathcal{Q}}
\def\scrr{\mathcal{R}}
\def\scrs{\mathcal{S}}
\def\scrt{\mathcal{T}}
\def\scrv{\mathcal{V}}
\def\scrw{\mathcal{W}}
\def\scrz{\mathcal{Z}}


\newenvironment{sarray}{
          \textfont0=\scriptfont0
          \scriptfont0=\scriptscriptfont0
          \textfont1=\scriptfont1
          \scriptfont1=\scriptscriptfont1
          \textfont2=\scriptfont2
          \scriptfont2=\scriptscriptfont2
          \textfont3=\scriptfont3
          \scriptfont3=\scriptscriptfont3
        \renewcommand{\arraystretch}{0.7}
        \begin{array}{l}}{\end{array}}

\newenvironment{scarray}{
          \textfont0=\scriptfont0
          \scriptfont0=\scriptscriptfont0
          \textfont1=\scriptfont1
          \scriptfont1=\scriptscriptfont1
          \textfont2=\scriptfont2
          \scriptfont2=\scriptscriptfont2
          \textfont3=\scriptfont3
          \scriptfont3=\scriptscriptfont3
        \renewcommand{\arraystretch}{0.7}
        \begin{array}{c}}{\end{array}}

\pagestyle{myheadings}
\markboth{{\rm Running head: RESPONSE ON POSITIVITY RATIOS}}{{\rm Running head: RESPONSE ON POSITIVITY RATIOS}}

\doublespace

Recently we (Brown, Sokal, \& Friedman, 2013)
debunked the widely-cited claim made by Fredrickson and Losada (2005)
that their use of a mathematical model drawn from nonlinear dynamics
(namely, the Lorenz equations from fluid dynamics)
provided theoretical support for the existence of a pair of
critical positivity-ratio values (2.9013 and 11.6346)
such that individuals whose ratios fall between these values
will ``flourish,'' whereas people whose ratios lie outside
this ideal range will ``languish.''\footnote{
   After the publication of Brown et al.\ (2013),
   Andr\'es Navas kindly drew our attention to his article (Navas, 2011)
   in which a very similar (though briefer) critique of Losada (1999)
   was made.
   [This footnote was unfortunately omitted from the published version
    of this article, due to space limitations.]
}
For lack of space in
our previous article,
we refrained from addressing, except in passing,
the question of whether there might be empirical evidence
for the existence of one or more critical positivity ratios
(``tipping points'').
In response to our critique,
Fredrickson and Losada (2013) withdrew their nonlinear-dynamics model,
but Fredrickson (2013) reaffirmed some claims concerning positivity ratios
on the basis of empirical studies.
We would therefore like to comment briefly on these claims
and the alleged supporting evidence.

The principal stated goal of Fredrickson (2013)
was ``to update the empirical evidence for the
value and nonlinearity of positivity ratios'' (p.~814).
Unfortunately,
that article is extremely unclear about which claims the author
has opted to renounce and which she has chosen to reaffirm.
This unclarity is, alas, compounded
by that in Fredrickson and Losada's (2013) subsequently issued
``correction'' to Fredrickson and Losada (2005),
which formally withdrew the ``modeling element'' and
``model-based predictions'' of this latter paper
but failed to identify which, if any, of Fredrickson's (2013)
arguments might also be impacted by this withdrawal.
In the present article, we refrain from second-guessing on this question
and instead concentrate on the published scientific literature:
that is, our critique focuses on Fredrickson (2013) as written.

\begin{figure}[t]
\centering
\begin{tabular}{c@{\hspace*{2cm}}c}
\includegraphics[width=0.28\textwidth]{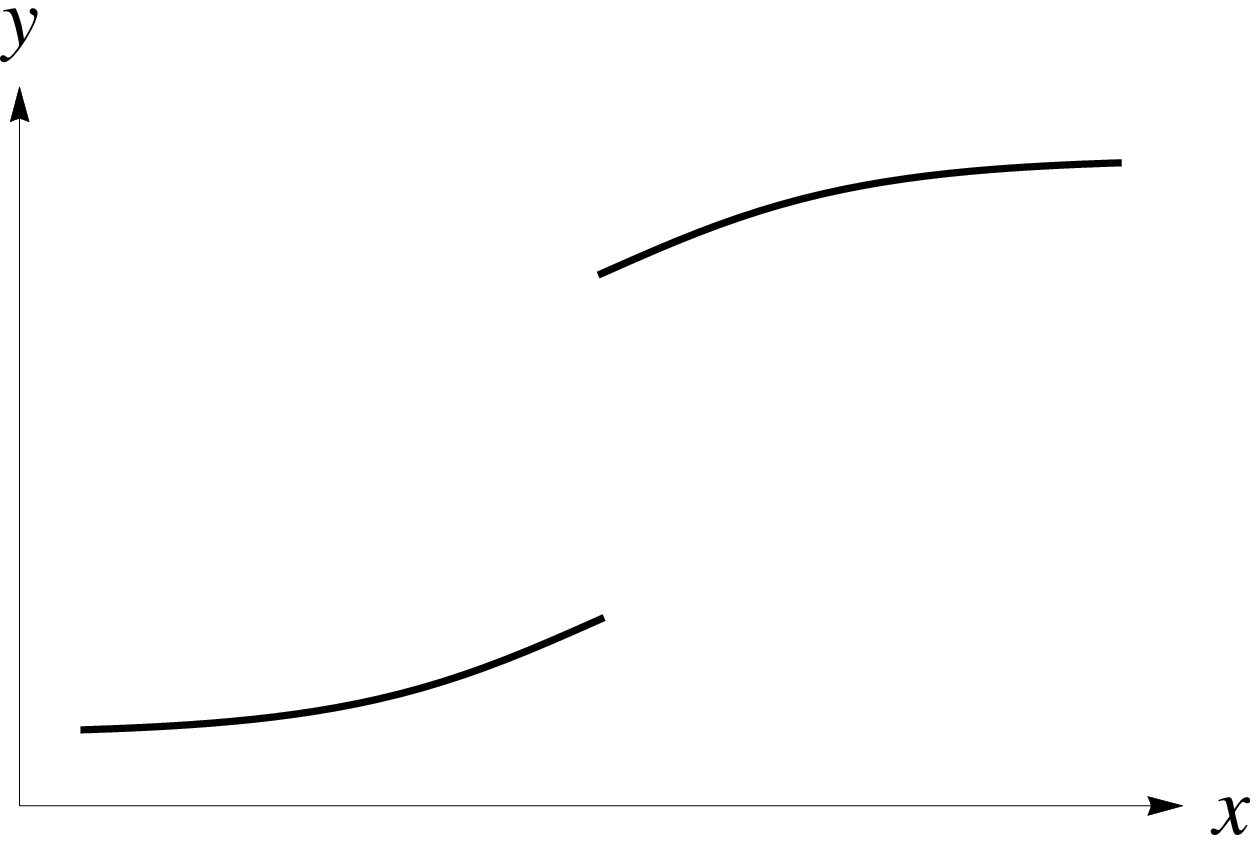} &
\includegraphics[width=0.28\textwidth]{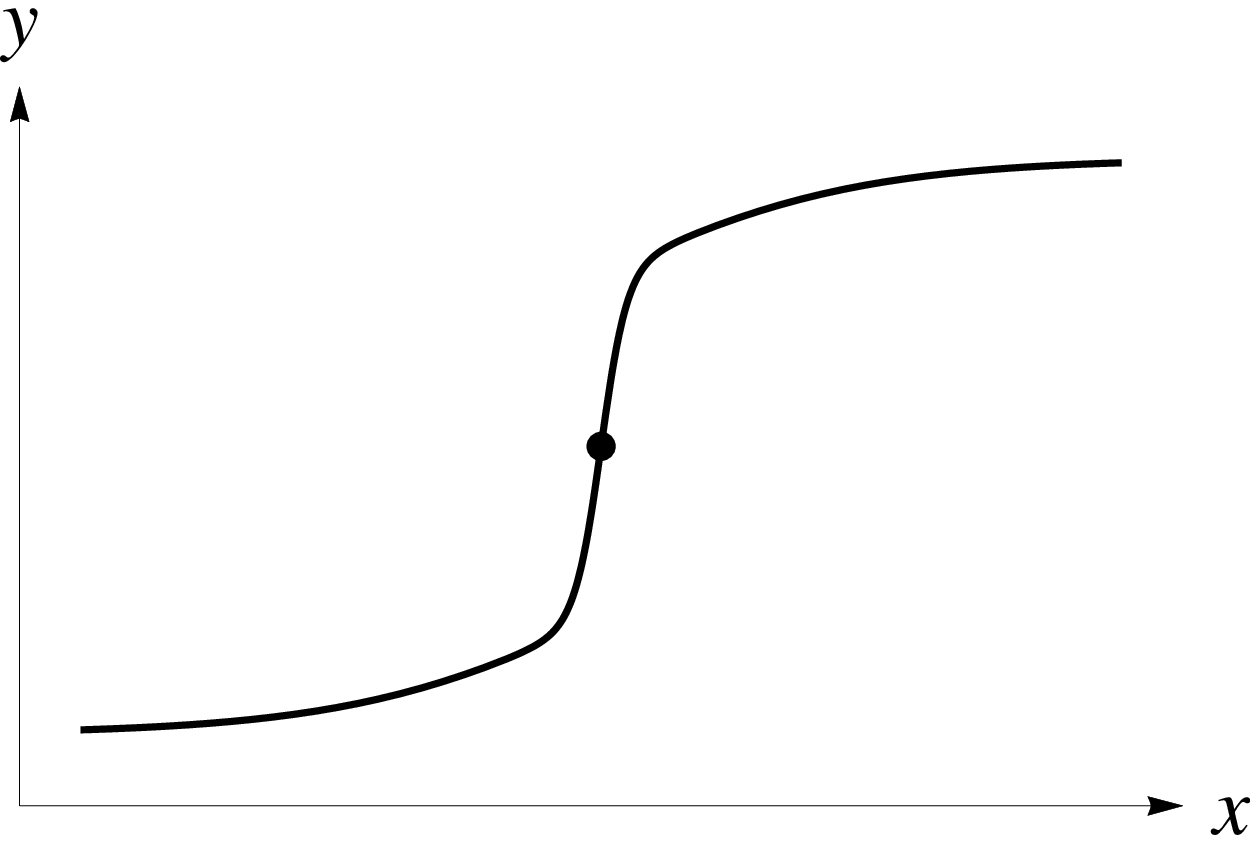} \\[1mm]
(a) & (b) \\[5mm]
\includegraphics[width=0.28\textwidth]{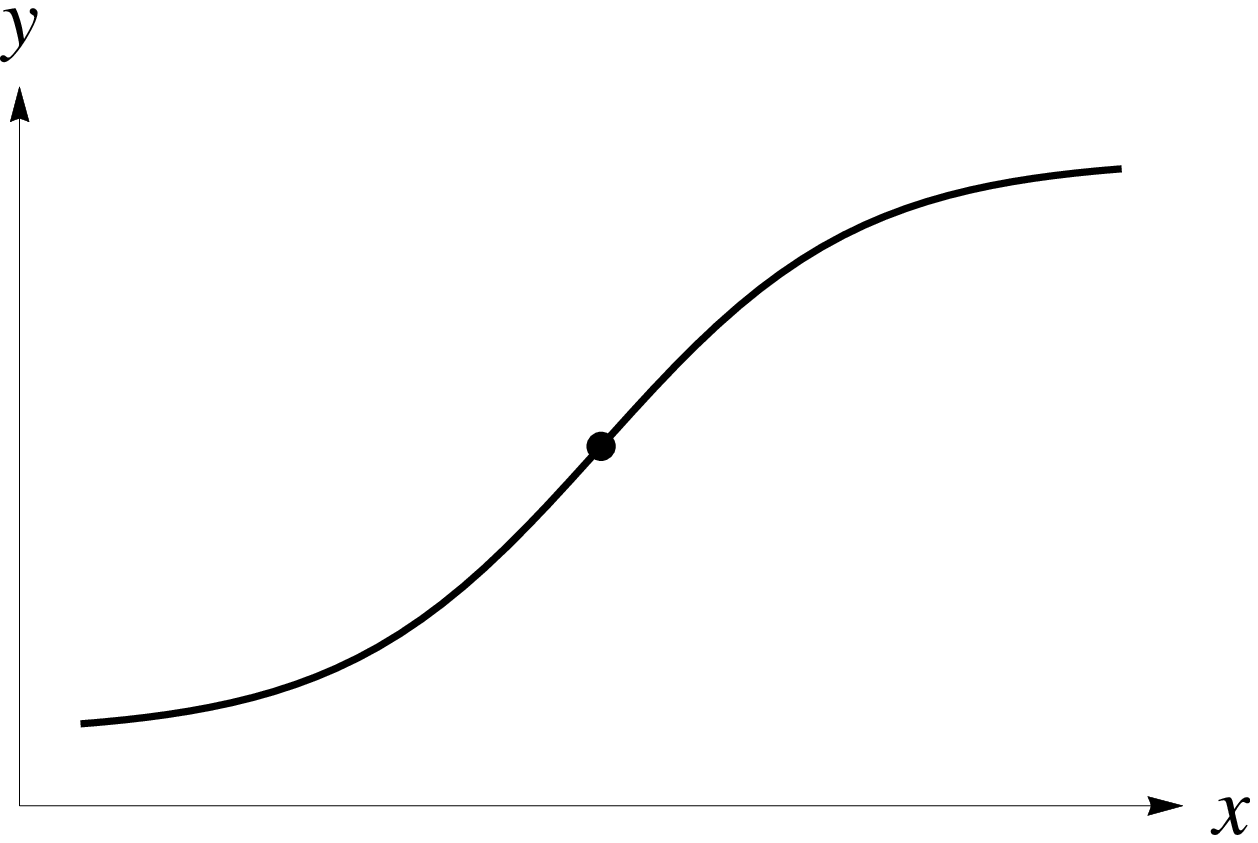} &
\includegraphics[width=0.28\textwidth]{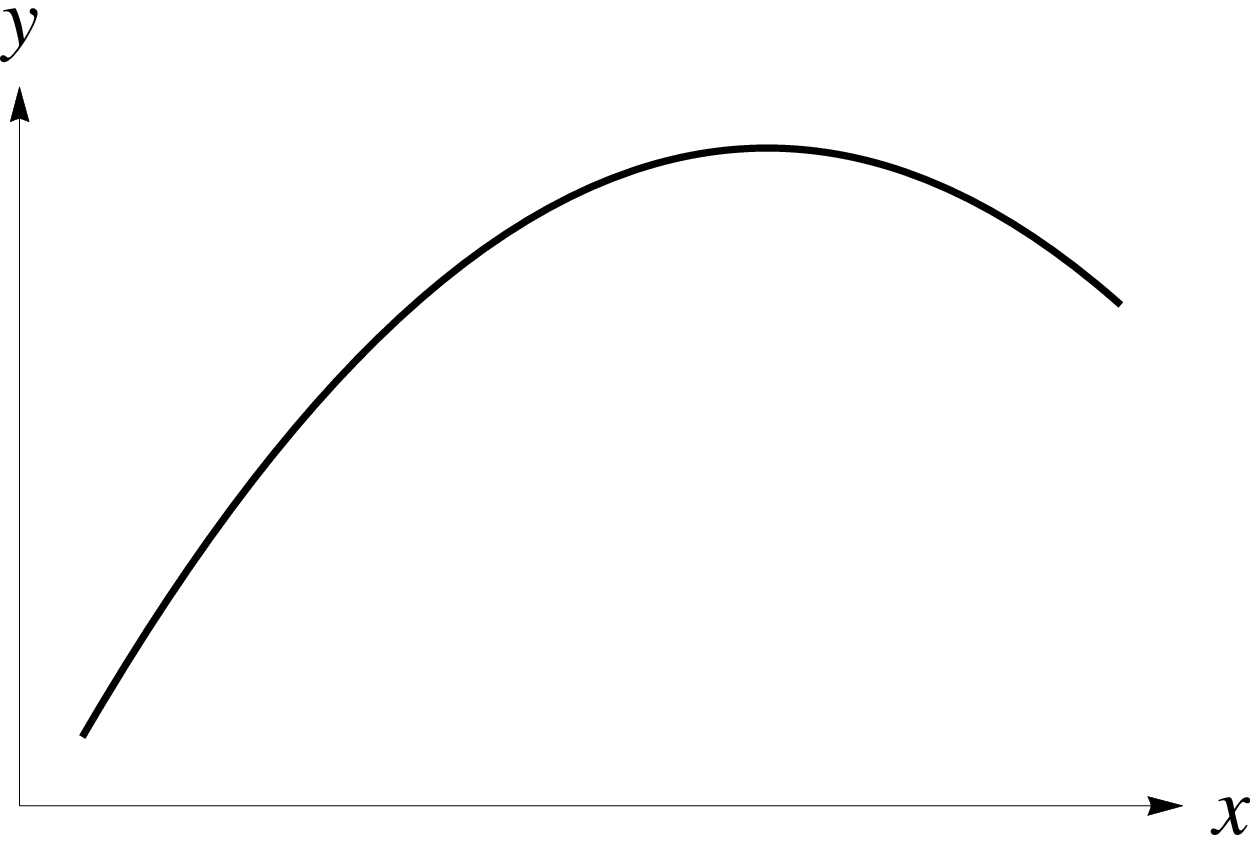} \\[1mm]
(c) & (d) \\
\end{tabular}
\caption{
   Some possible nonlinear behaviors of ``degree of flourishing'' ($y$)
   as a function of the positivity ratio ($x$):
   (a) Discontinuous phase transition (``tipping point'').
   (b) Rapid change.
   (c) Gradual change with inflection point.
   (d) Inverted-U.
   In (b, c) the inflection point is indicated with a dot.
 \label{fig1}
}
\end{figure}

To clarify what is at stake,
consider the following sequence of successively weaker claims
for the behavior of ``degree of flourishing''
as a function of the positivity ratio
(see
also
Figure~\ref{fig1}):\footnote{
   For simplicity we consider here only the claims
   concerning the alleged lower critical positivity ratio (2.9013).
   Similar considerations apply also to the claims
   concerning the alleged upper critical positivity ratio (11.6346).
}
\vspace*{-2mm}
\begin{enumerate}
\itemsep1pt \parskip0pt \parsep0pt
  \item There is a discontinuous phase transition (``tipping point'')
       exactly at 2.9013.  
  \item There is a discontinuous phase transition 
       somewhere around 3.  
  \item There is a rapid change somewhere around 3.   
  \item There is an inflection point (separating convexity from concavity)
            somewhere around~3.   
  \item There is an inflection point (separating convexity from concavity)
            somewhere.   
  \item There is some nonlinearity somewhere.
\end{enumerate}
Fredrickson and Losada (2005) made claim \#1;
Fredrickson (2009, especially Chapter~7) reaffirmed claim \#1
but noted that, because of ``impurities'' and measurement imprecision
(p.~129),
the data might look in practice more like claim \#2 or \#3.
What does Fredrickson (2013) assert?
Alas, this is shrouded in confusion.
The only reason for even considering
the possibility of a discontinuous phase transition
was the nonlinear-dynamics model based on the Lorenz equations,
which we (Brown et al., 2013) have shown to be entirely groundless
and which Fredrickson and Losada (2013) have now withdrawn.
Nevertheless, Fredrickson (2013, p.~819)
insisted that
such a transition
remains a viable possibility:
``Whether the outcomes associated with positivity
ratios show discontinuity and obey one or more
specific change points, however, merits further test.''
Indeed, Fredrickson (2013, p.~818)
continued to insist that even
the original
claim \#1 remains a viable possibility:
``The question \ldots\ is whether positivity ratios obey one or more
critical tipping points, and if so, whether those critical tipping points
coincide with the ones identified by Losada's mathematical work
for all individuals, samples, and subgroups.
Clearly, these questions merit further test.''
[This latter contention is somewhat puzzling in the light of our
demonstration (Brown et al., 2013, pp.~811--812), not refuted
or even addressed by Fredrickson (2013), that the mathematical
model of Fredrickson and Losada (2005) does not make {\em any}\/
definite predictions for the ``critical tipping points,''
since those depend on completely arbitrary choices of the
constants $\sigma$, $b$, and $i$.]
But despite these insistences that a discontinuous phase transition
remains a viable possibility, Fredrickson (2013) did not present
any evidence that
such a discontinuity occurs or is even plausible.
Rather, in summarizing recent empirical work
that in her view supports the existence of nonlinearity,
she appeared to be arguing for claim \#4, \#5 or \#6
(it is not clear which).
We will address this empirical evidence below.

Fredrickson (2013) also did not make clear whether she was continuing
to advocate the use of {\em nonlinear-dynamics models}\/ as in
Fredrickson and Losada (2005),
or merely arguing for the investigation of
{\em nonlinear relations between variables}\/
along the lines just discussed.
These are two quite different matters,
but Fredrickson sometimes wrote as if they were interchangeable.
For example:
\begin{quote}
\small
Thus, while Brown and colleagues (2013) urged caution
in the use of nonlinear dynamics, I will show that the available evidence
makes clear that researchers investigating affective phenomena
need to recognize and mathematically address growing evidence
for nonlinearity in their data.
\quad (Fredrickson, 2013, pp.~816--817)
\end{quote}
Though her empirical evidence
concerned only the question of nonlinear relations,
at some points in her article she forcefully advocated the use
of (unspecified) nonlinear-dynamics models.
For instance, while conceding that
``the nonlinearity evident in human emotion systems
may not be best modeled by the specific set of differential equations
that Losada proposed,''
she nevertheless remained
``convinced of the need to identify and test mathematical
and statistical models that are sensitive to nonlinear,
recursive, and dynamic effects'' (p.~817).
She further argued that since
``human emotions are clearly dynamic, multicomponent systems
that show self-sustaining upward and downward spirals sensitive
to changing circumstances,''
it therefore follows (according to her) that
``system dynamics, network analysis, agent-based modeling,
and other systems science approaches are likely to become ever more relevant
to affective science and positive psychology'' (p.~820).
Indeed, Fredrickson (2013) was not yet prepared to abandon the attempt to model
the time evolution of human emotions using the Lorenz equations:
``Whether the Lorenz equations --- the nonlinear dynamic
model we'd adopted --- and the model estimation
technique that Losada utilized can be fruitfully applied to
understanding the impact of particular positivity ratios merits
renewed and rigorous inquiry'' (p.~814).\footnote{
   In fact, there is no evidence
   in any of the three published articles
   (Losada, 1999; Losada \& Heaphy, 2004; Fredrickson \& Losada, 2005)
   that Losada or his colleagues utilized {\em any}\/
   ``model estimation technique''
   (see Brown et al., 2013, pp.~806--808, 811).
}
Let us stress that we have no objection in principle to such models,
but simply insist that valid applications of differential equations
must meet the criteria VA1--VA5 set forth in our article
(Brown et al., 2013, pp.~804--805).

Let us now turn to the empirical evidence bearing on claims \#1--6,
which should in turn be compared against two alternative scenarios:
\begin{itemize}
  \item[7.] There is a positive (but substantially linear) correlation
       between the positivity ratio and ``degree of flourishing.''  \\[-10mm]
  \item[8.] There is no correlation
       between the positivity ratio and ``degree of flourishing.''
\end{itemize}

Fredrickson and Losada (2005) studied two samples of college students
and purported to find empirical evidence in favor of claim \#1.
Unfortunately, their study design and method of analysis
were such that {\em no data whatsoever}\/ from these could provide
{\em any}\/ evidence for {\em any}\/ nonlinearity
(i.e., even the weakest claim \#6) ---
because the information that might provide this evidence
was discarded at an early stage, when participants were dichotomized,
based on their scores on a 33-item questionnaire
of positive psychological and social functioning,
as ``flourishing'' or ``nonflourishing.''
Although the subsequent finding that the mean positivity ratio
in the ``flourishing'' group was significantly higher than in the
``nonflourishing'' group provides
some evidence of a positive correlation between
the positivity ratio and ``degree of flourishing''
(i.e., evidence against scenario \#8)\footnote{
   Let us mention in passing that the reported result for Sample 2
   [$t(99) = 1.62$] yields $p = 0.0542$ (one-tailed);
   it is {\em not}\/ correct to claim statistical significance
   by rounding this down to $0.05$ (p.~684)!
   Also, it is debatable whether it is valid to use a one-tailed
   test to determine whether the means of two groups are different,
   absent strong theoretical or logical reasons,
   {\em independent of the theory under test}\/,
   for believing that the first mean (for instance)
   must necessarily be greater than or equal to the second.
},
this result
is perfectly compatible both with a linear correlation (scenario \#7)
and with a nonlinear correlation (claim \#6);
the dichotomized data are totally incapable of distinguishing these
two alternatives.
The fact that ``these mean ratios flanked the 2.9 ratio,''
which Fredrickson and Losada considered ``critical to our hypothesis''
(p.~684)
--- an argument reiterated by Fredrickson (2013, pp.~817--818) ---
is in fact utterly irrelevant:
it provides no evidence even for claim \#6, much less for claim \#1.
The same comments apply, {\em mutatis mutandis}\/,
to Waugh and Fredrickson (2006) and Diehl, Hay, and Berg (2011),
which were cited by Fredrickson (2013) in support of
her claims of nonlinearity.

The correct way to test for nonlinearity is to form the scatter plot
between the positivity ratio ($x$) and some quantitative
({\em not}\/ artificially dichotomized or trichotomized)
measure of ``flourishing'' ($y$),
and to perform various statistical tests on it;
in particular, one can try to estimate the regression curve $Y(x)$
and to test whether there are statistically significant deviations
from linearity (and if so, of what magnitude and what form).
Such an approach was taken by
Rego, Sousa, Marques, and Cunha (2012)
and by Shrira et al.\ (2011);
both of these studies found weak evidence,
to be discussed below,
for some form of nonlinearity (i.e., claim \#6).
None of the studies cited by Fredrickson (2013)
--- or any others we are aware of ---
contain any evidence for an inflection point (claim \#5),
much less for a ``tipping point''
(claim \#1 or \#2).
Indeed, the data of Rego et al.\ (2012) and Shrira et al.\ (2011)
constitute strong evidence {\em against}\/ the hypothesis of a
``tipping point''.
See, for instance, the data of Rego et al.\ 
displayed in Figure~\ref{fig2} below:
there is no indication whatsoever
of any discontinuous behavior (claim \#2),
any rapid change (claim \#3),
or even any inflection point (claim \#5).

But more is true:
Fredrickson and Losada (2005)'s {\em own}\/ raw data
almost certainly constitute strong evidence {\em against}\/
their hypothesis of a ``tipping point'' at 2.9013.
(We say ``almost certainly'' because Professor~Fredrickson
has informed
us that the raw data are no longer available;
 we must therefore proceed, {\em faute de mieux}\/,
 on the basis of the published summary statistics.)
To understand the logic here,
recall first that according to the Fredrickson--Losada hypothesis (claim \#1),
{\em all}\/ ``flourishing'' individuals should have positivity ratios
above 2.9013, while {\em all}\/ ``nonflourishers''
should have positivity ratios below 2.9013.
Let us even be charitable and allow for
``impurities'' and measurement imprecision
(Fredrickson, 2009, p.~129)
by replacing ``all'' by ``almost all'' in this hypothesis.
Are Fredrickson and Losada's data consistent with the hypothesis,
thus modified?
Unfortunately we don't know for sure, because Fredrickson and Losada (2005)
failed to report {\em anything}\/ about the positivity-ratio distributions
for the ``flourishing'' and ``nonflourishing'' groups,
other than their means.
But we can obtain a rough estimate by reverse-engineering their statistics,
under the simplifying hypothesis that the ``flourishing'' and
``nonflourishing'' groups have the same standard deviation $\sigma$;
this allows us to apply, in reverse, the formula for the $t$-test
assuming equal variance but unequal sample sizes
(Koopmans, 1981, p.~302).\footnote{
   The $t$-test for samples of sizes $n_1$ and $n_2$,
   with sample means $\bar{X}_1$ and $\bar{X}_2$,
   taken from populations having the same standard deviation $\sigma$,
   is
   $$ t  \;=\;  {\bar{X}_1 - \bar{X}_2
                 \over
                 s \, \sqrt{\displaystyle {1 \over n_1} + {1 \over n_2}}
                }
   $$
   where $s$ is an
estimator of the common standard deviation $\sigma$.
   We used this equation in reverse to compute~$s$.
   [This footnote was omitted from the published version
    of this article, due to space limitations.]
}
Then their data for Sample 1
($n_1 = 36$, $n_2 = 51$, $\bar{X}_1 = 3.2$, $\bar{X}_2 = 2.3$, $t = 2.32$)
imply $\sigma \approx 1.78$,
while their data for Sample 2
($n_1 = 9$, $n_2 = 92$, $\bar{X}_1 = 3.4$, $\bar{X}_2 = 2.1$, $t = 1.62$)
imply $\sigma \approx 2.30$.
On the other hand, {\em every}\/ probability distribution
supported on the interval $[0,M]$ and having mean $\mu$
has a variance $\sigma^2 \le \mu (M-\mu)$;
hence $M \ge \mu + (\sigma^2/\mu)$.
Applying this to the ``nonflourishing'' groups in Samples 1 and 2,
we obtain $M \gtapprox 3.68$ and $M \gtapprox 4.61$, respectively ---
values far in excess of 2.9013.
It is therefore {\em inevitable}\/ that the ``nonflourishing'' groups
contained large numbers of individuals with positivity ratios
in excess of 2.9013;
assuming a normal distribution
we estimate that approximately 36\% of ``nonflourishers''
had positivity ratios above 2.9013.

\begin{figure}[t]
\centering
\begin{tabular}{c@{\hspace*{1cm}}c}
\hspace*{-4mm}
\includegraphics[width=0.46\textwidth]{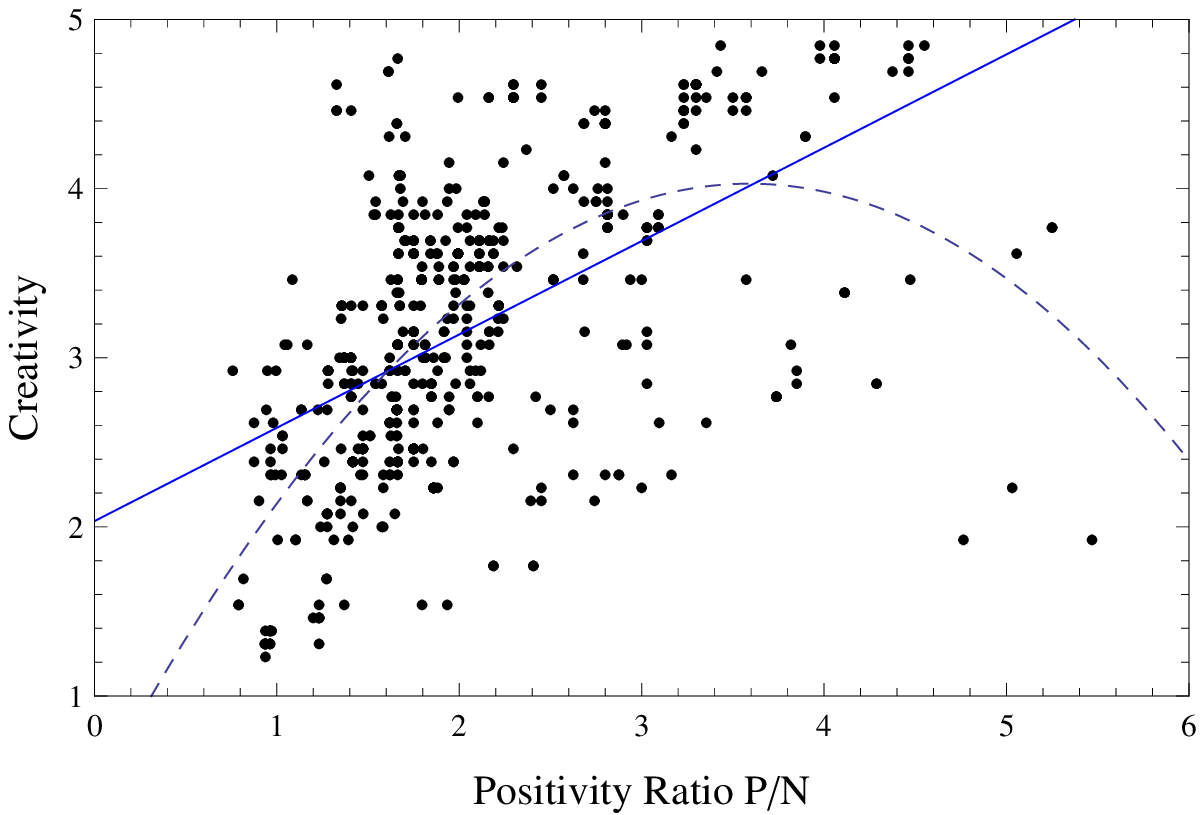} &
\includegraphics[width=0.46\textwidth]{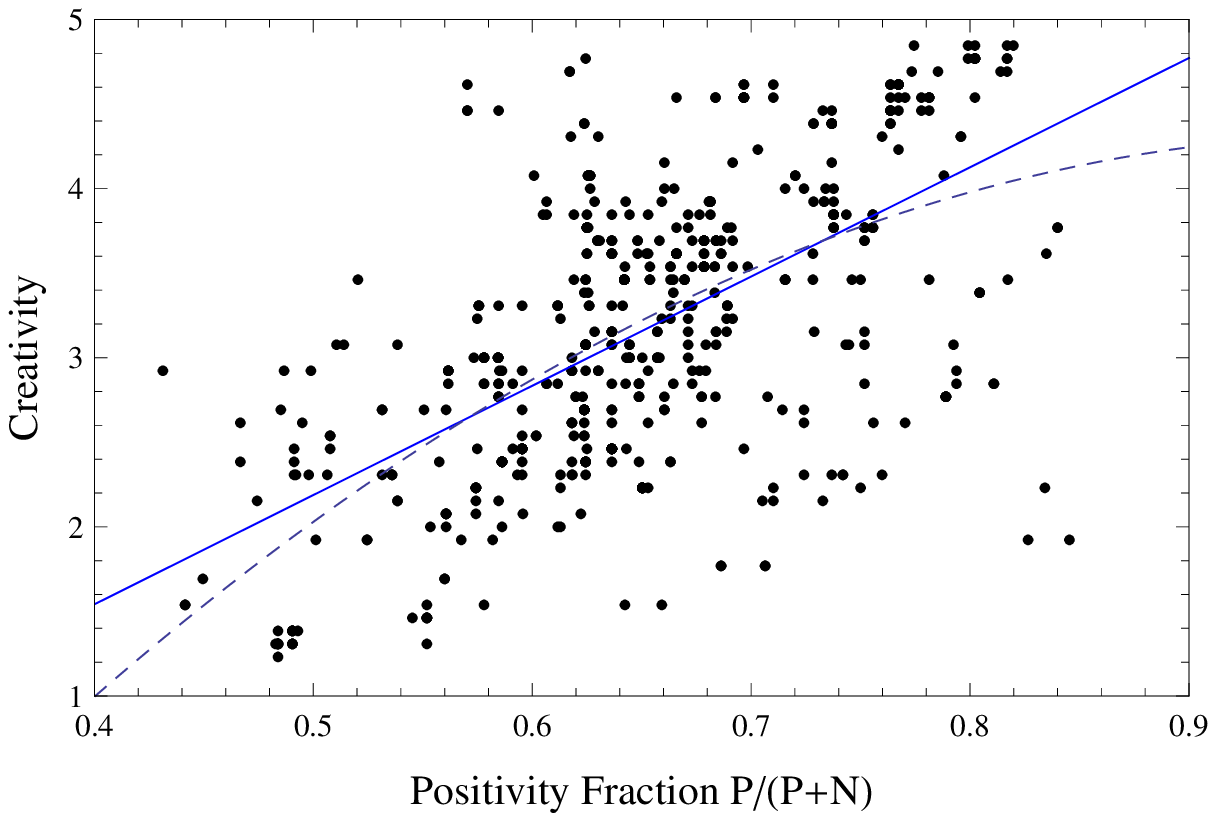} \\
(a) & (b) \\
\end{tabular}
\caption{
   Linear and quadratic fits to the data of Rego et al.\ (2012),
   with the independent variable being
   (a) the positivity ratio $P/N$ or
   (b) the positivity fraction $P/(P+N)$.
 \label{fig2}
}
\end{figure}

So let us forget about ``tipping points'' (claims \#1 and \#2)
and go back to the weak evidence for concave nonlinearity (claim \#6)
found by Rego et al.\ (2012) and Shrira et al.\ (2011).
It turns out that this nonlinearity
can be almost entirely removed by a simple and well-motivated
change of independent variable.
Notice first that the
regression curve of
``degree of flourishing'' {\em cannot possibly}\/
be a perfectly linear function of the positivity ratio,
for the simple reason that the ``degree of flourishing'' is bounded
(for instance, Rego et al.'s measure of ``creativity'' runs from 1 to~5)
while the positivity ratio can in principle become arbitrarily large.
This suggests that a much better choice of independent variable,
in place of the positivity ratio $P/N$, would be the
quantity
$P/(P+N)$,
which runs from 0 to 1
(we might call it the {\em positivity fraction}\/);
it contains exactly the same information as $P/N$
but in a form that avoids creating superfluous nonlinearities.\footnote{
   As an added advantage, $P/(P+N)$ also avoids
   the undesirable arithmetic effects that can occur
   when participants report very few (or zero) negative emotions
   during the specified time interval.
   Fredrickson and Losada (2005, p.~681, footnote 1)
   themselves observed that the conversion between $P/N$ and $P/(P+N)$
   is simply an algebraic transformation.
}
In Figure~\ref{fig2} we compare the fits for the data of Rego et al.\ 
using $P/N$ versus $P/(P+N)$ as the independent variable.
In the latter variable, the quadratic coefficient is still
statistically significant ($b_2 = -9.6 \pm 3.1$)
but its effect is quantitatively small;
the linear and quadratic fits differ only very slightly
in the region where most of the data points lie.
Broadly similar considerations apply to the data of Shrira et al.\ (2011).

There is, however, a more fundamental problem with all these studies
(or rather, with their interpretation).
The idea of searching for
{\em the}\/
relation
between the positivity ratio and ``flourishing''
--- and in particular of testing whether it is nonlinear ---
presupposes that the question is well-posed.  But is it?
Of course one can seek to estimate the regression curve $Y(x)$
in any given population.
But what reason is there to believe that different populations
will produce the {\em same}\/ regression curve?
Indeed, what reason is there to believe that even {\em qualitative}\/
features of that curve --- such as its monotonicities and convexities ---
will be universal?
Perhaps
(to take just one possibility)
higher positivity ratios produce higher ``well-being''
for populations living in generally harmonious and prosperous environments
(for instance, American college students)
but 
produce lower ``well-being'' for populations exposed to
danger and predation (for instance, during
a civil war).
From this perspective it is hardly surprising that the studies
cited by Fredrickson (2013) produced such wildly divergent results:
   Rego et al.\ (2012) and Shrira et al.\ (2011) found weak evidence
   that the ``degree of flourishing'' ceased to increase significantly,
   and might conceivably even decrease slightly,
   when positivity ratios exceed approximately 3,
   whereas Diehl et al.\ (2011)
   found monotonically increasing levels of mental health
   (which was, alas, trichotomized
    as ``languishing,'' ``moderately mentally healthy,'' or ``flourishing'')
   right up to the highest mean positivity ratios observed in their study
   --- which were 4.3 (young adults),
   14.8 (middle-aged adults) and 26.7 (older adults).

Fredrickson (2009, p.~134; 2013, p.~819) described what she called the
``Now you see it, now you don't'' effect of positivity,
which appears to be ``playing an elaborate shell game with scientists'':
attempts to measure the effects of positive emotions
sometimes ``find no effects whatsoever''
because,
according to her,
these effects become reliably detectable only when
positivity ratios exceed the ``tipping point'' value.
We wish to put forward an alternative hypothesis:
namely, that when no effect is found,
this is probably because no effect exists
(or it is too weak to be detected by the given study).
This alternative hypothesis is not only simpler
than the ``shell game'' scenario,
but has the added advantage of not relying on the hypothetical existence
of a ``tipping point,'' for which, as we have shown,
there is at~present no evidence whatsoever
(and indeed significant evidence against).

One of the concluding remarks in our article (Brown et al., 2013, p.~813)
addressed the need for the use of differential-equation models
in the natural or social sciences
to be duly justified by theoretical arguments and/or empirical evidence.
Perhaps we should have widened that remark to take in all
mathematical modeling
of observed phenomena.
To have any predictive value, mathematical models
must be rigorously specified and shown to accurately reflect reality.
The mere invocation of buzzwords such as
``nonlinear,'' ``recursive,'' and ``dynamic'' (Fredrickson, 2013, p.~817)
carries no weight in responsible scholarship, despite their
romantic appeal.

\bigskip
\bigskip

{\small
We wish to thank Arm\'enio Rego and Amit Shrira for providing us,
respectively, the data from Rego et al.\ (2012)
and from Study~1 of Shrira et al.\ (2011).
We are also extremely grateful to Andrew Gelman and Carol Nickerson
for helpful comments on an early draft of this manuscript.
The authors, of course, bear full and sole responsibility
for the content of this article.

Correspondence concerning this article should be addressed to
Alan D.~Sokal, Department of Physics, New York University,
4 Washington Place, New York, NY 10003.
E-mail: sokal@nyu.edu
}

\bigskip
\bigskip

\begin{center} {\bf References} \end{center}

\small

\begin{hangparas}{.5in}{1}

Brown, N. J. L., Sokal, A. D., \& Friedman, H. L. (2013).
The complex dynamics of wishful thinking: The critical positivity ratio.
{\em American Psychologist, 68}\/, 801--813.
doi:10.1037/a0032850

Diehl, M., Hay, E. L., \& Berg, K. M. (2011).
The ratio between positive and negative affect and flourishing mental health
across adulthood.
{\em Aging \& Mental Health, 15}\/, 882--893.
doi:10.1080/13607863.2011.569488

Fredrickson, B. L. (2009).
{\em Positivity: Groundbreaking research reveals how to embrace
  the hidden strength of positive emotions, overcome negativity, and thrive}\/.
New York, NY: Crown.
Published in paperback as 
{\em Positivity: Top-notch research reveals the 3-to-1 ratio
that will change your life}\/.

Fredrickson, B. L. (2013).
Updated thinking on positivity ratios.
{\em American Psychologist, 68}\/, 814--822.
doi:10.1037/a0033584

Fredrickson, B. L., \& Losada, M. F. (2005).
Positive affect and the complex dynamics of human flourishing.
{\em American Psychologist, 60}\/, 678--686.
doi:10.1037/0003-066X.60.7.678

Fredrickson, B. L., \& Losada, M. F. (2013).
Correction to Fredrickson and Losada (2005).
{\em American Psychologist, 68}\/, 822.
doi:10.1037/a0034435

Koopmans, L. H.  (1981).
{\em An introduction to contemporary statistics.}\/
Boston, MA: Duxbury Press.

Losada, M. (1999).  The complex dynamics of high performance teams.
{\em Mathematical and Computer Modelling, 30}\/(9--10), 179--192.
doi:10.1016/S0895-7177(99)00189-2

Losada, M., \& Heaphy, E. (2004).
The role of positivity and connectivity in the performance of business teams:
A nonlinear dynamics model.
{\em American Behavioral Scientist, 47}, 740--765.
doi:10.1177/0002764203260208

Navas, A.  (2011).  Un cas d'inconscience (?).
{\em Images des Math\'ematiques}\/, CNRS.
Retrieved from
http://images.math.cnrs.fr/Un-cas-d-inconscience,1004.html

Rego, A., Sousa, F., Marques, C., \& Cunha, M. P. (2012).
Optimism predicting employees' creativity:
The mediating role of positive affect and the positivity ratio.
{\em European Journal of Work and Organizational Psychology, 21}\/, 244--270.
doi:10.1080/1359432X.2010.550679

Shrira, A., Palgi, Y., Wolf, J. J., Haber, Y., Goldray, O.,
Shacham-Shmueli, E., \& Ben-Ezra, M. (2011).
The positivity ratio and functioning under stress.
{\em Stress and Health, 27}\/, 265--271.
doi:10.1002/smi.1349

Waugh, C. E., \& Fredrickson, B. L. (2006).
Nice to know you: Positive emotions, self--other overlap,
and complex understanding in the formation of a new relationship.
{\em Journal of Positive Psychology, 1}\/, 93--106.
doi:10.1080/17439760500510569

\end{hangparas}

\end{document}